# Temperature Dependence of the Refractive Index for AlAsGaSb

HELENA JANOWSKA,[1,*] WOJCIECH CHARASZKIEWICZ,[1] MAJA WASILUK,[1] MARKUS PEIL,[2] TEEMU TASKINEN,[2] JOONAS HILSKA,[2] ABHIROOP CHELLU,[2] TEEMU HAKKARAINEN,[2,3] AND ANNA MUSIAŁ[1]

[1]*Laboratory for Optical Spectroscopy of Nanostructures, Department of Experimental Physics, Faculty of Fundamental Problems of Technology, Wrocław University od Science and Technology, Wybrzeże Wyspiańskiego 27, 50-370 Wrocław, Poland*
[2]*Optoelectronics Research Centre, Physics Unit, Korkeakoulunkatu 10, 33720 Tampere, Finland*
[3]*Tampere Institute for Advanced Study, Tampere University, Ratapihankatu 55, FI-33014 Tampere, Finland*
**e-mail address of corresponding author: 268275@student.pwr.edu.pl*

**Abstract:** Accurate design and optimization of photonic multilayer structures like distributed Bragg reflectors (DBRs) require precise knowledge of material optical constants, particularly the temperature dependence of the refractive index. While these parameters are well established for widely used semiconductors, for emerging materials such as antimonides they are often limited to room-temperature data, especially for new spectral ranges of interest. Antimonide compounds, in particular GaSb-based alloys, are promising for quantum photonics applications. In this work, we investigated DBRs lattice-matched to GaSb and designed for operation in the third telecommunication window. Reflectivity spectra were measured in the temperature range from 11.5 K to 300 K, and then fitted using the transfer matrix method (TMM), combined with a dedicated recursive numerical fitting algorithm. Initial parameters included layer thicknesses determined by scanning electron microscopy (SEM) and literature values of refractive indices at room temperature. This approach enabled extraction of the temperature-dependent refractive indices of two AlGaAsSb alloys suitable for forming DBR mirrors for 1.5 μm wavelengths. The obtained results provide essential input for reliable DBR design, ensuring proper stopband positioning and high reflectivity under cryogenic operating conditions required for efficient quantum emitter performance.

## 1. Introduction

The refractive index of semiconductor materials is one of the most important parameters when it comes to designing photonic structures and devices. Proper usage of materials based on their refractive indices provides us with ability to exploit such structures as, e.g., DBRs or Circular Bragg Grating Cavities (CBGC), Fabry-Perot defect cavities, and open microcavities. These have been successfully used to strongly enhance signal collection efficiency or to optimize the parameters of non-classical light sources [1-6]. However, achieving high efficiency of DBRs is challenging due to the necessity of using many alternating material layers, which in the case of epitaxial growth, leads to creating strain that influences quality of each layer due to lattice mismatch between used materials. For optical cavities refractive indices of materials determine light confinement, and the final performance of the light source is very sensitive to its values, especially in the case of high quality and low mode volume cavities.

Difficulty in designing photonic structures for non-classical light sources, especially in the telecommunication spectral range, comes from necessity of operating in cryogenic temperature regime if best performance is to be achieved. Whereas standard experimental techniques are applicable at room temperature and the refractive indices values are typically not well-known for cryogenic temperatures. This is especially the case for materials that have not yet been used in this context. One of such emerging group of materials, that has recently been used for achieving single-photon emission in the third telecommunication window are Sb-based alloys using local droplet etching technique [7-9] – growth method so far most successful in realization of non-classical light sources with application-relevant parameters at 780 nm [10]. The interest in these particular materials comes from their high refractive index contrast [11] and low lattice mismatch with different alloy composition allowing for easier growth process, for optical structures and enhancing their efficiency. Moreover, crystal structure of these materials allows for strain relief convenient for growth on more accessible substrates like GaAs or Si [11,12], which is useful for implementing such structures in more complex optoelectronic devices and interfacing electronic and photonic integrated circuits.

Although the refractive index of GaSb-based alloys in room temperature has been measured before with ellipsometry [14] or interferometry [15], the temperature dependence of the material properties has not been yet studied experimentally. However, even room temperature results show huge variation in the determined refractive indices [16]. Previous attempts to measure this dependance for GaSb lack experimental data for the most crucial temperature and spectral regimes when it comes to quantum communication applications [17].

In this article we present the temperature dependence of refractive indices of Sb based alloys in the range of the third telecommunication window which presents great potential in the field of optoelectronic devices and non-classical light sources exploiting quantum electrodynamics and quantum optics for quantum communication. We acquire refractive index values by measuring reflectivity spectra of DBR structures consisting of AlGaAsSb alloys and fitting them using transfer matrix method (TMM) [18].

## 2. Investigated structures

The layer structure of the two DBR samples A and B investigated in our study are presented in Fig. 1. Both structures were grown via molecular beam epitaxy on n-GaSb(100) substrates with two different $Al_{0.32}GaAsSb$/$Al_{0.9}GaAsSb$ DBR structures. For both samples, the DBRs were grown at 500°C growth temperature measured via pyrometry and with moderate V/III flux ratios between 1.5–2.0. In each of the DBR layer pairs, the As and Sb fluxes were adjusted to lattice match the $Al_xGa_{1-x}As_ySb_{1-y}$ layers to the GaSb substrate, i.e. with compositional relationship of $y = 0.08 \cdot x$ [19]. The layer structure for sample A consists of a simple 6-pair DBR-stack, whereas sample B consists of optical 1λ-cavity with 2-pair DBR at the top and 3-pair DBR at the bottom. The λ-cavity of sample B is 407.1 nm thick and consists, among other layers, of an inner 0.5λ As-free $Al_{0.3}GaSb$ layer. Sample B is also capped with an additional 5 nm GaSb capping layer to protect from atmospheric oxidation. For sample A and B, the DBR design wavelengths were 1550 nm and 1500 nm, respectively. The DBR structures are made with pairs of high refractive index $Al_{0.9}Ga_{0.1}As_{0.06}Sb_{0.94}$ (120.6±2.3 nm for sample A and 112.6±2.5 nm for sample B, respectively) layer and low refractive index $Al_{0.32}Ga_{0.68}As_{0.02}Sb_{0.98}$ (104.6±1.8 nm for sample A and 102.6±2.0 nm for sample B, respectively) layer. These values are determined based on the cross-sectional Scanning Electron Microscopy (SEM) measurements shown in Fig. 2. Layer thicknesses were estimated from raw high-magnification secondary electron detector image (Fig. 2b) with Prewitt edge-detection in Matlab. The thicknesses reported here correspond to the average value of the constituent layers with standard deviation as the measurement uncertainty.

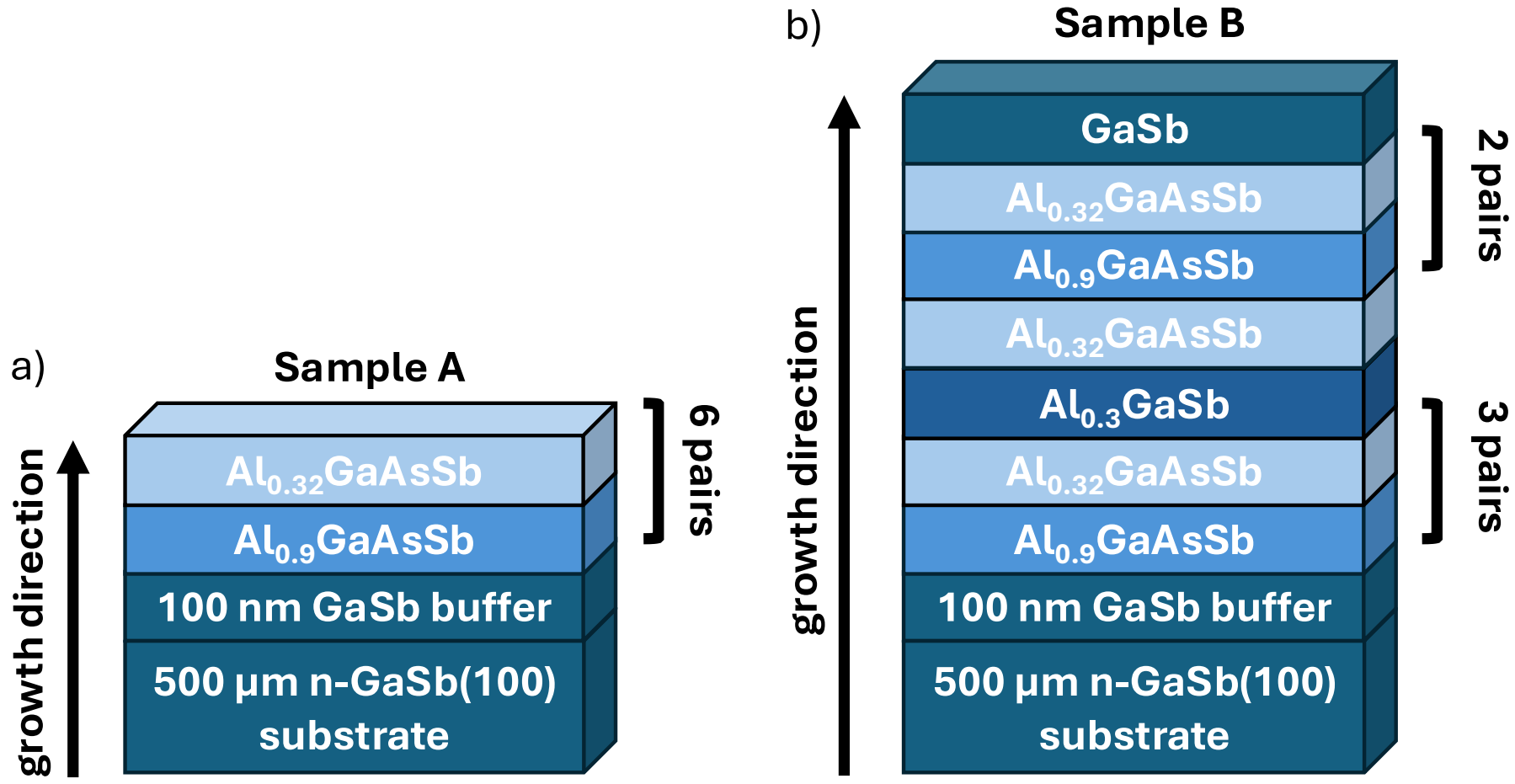

Fig. 1. The scheme of (a) Sample A (DBR only), and (b) Sample B (optical cavity).

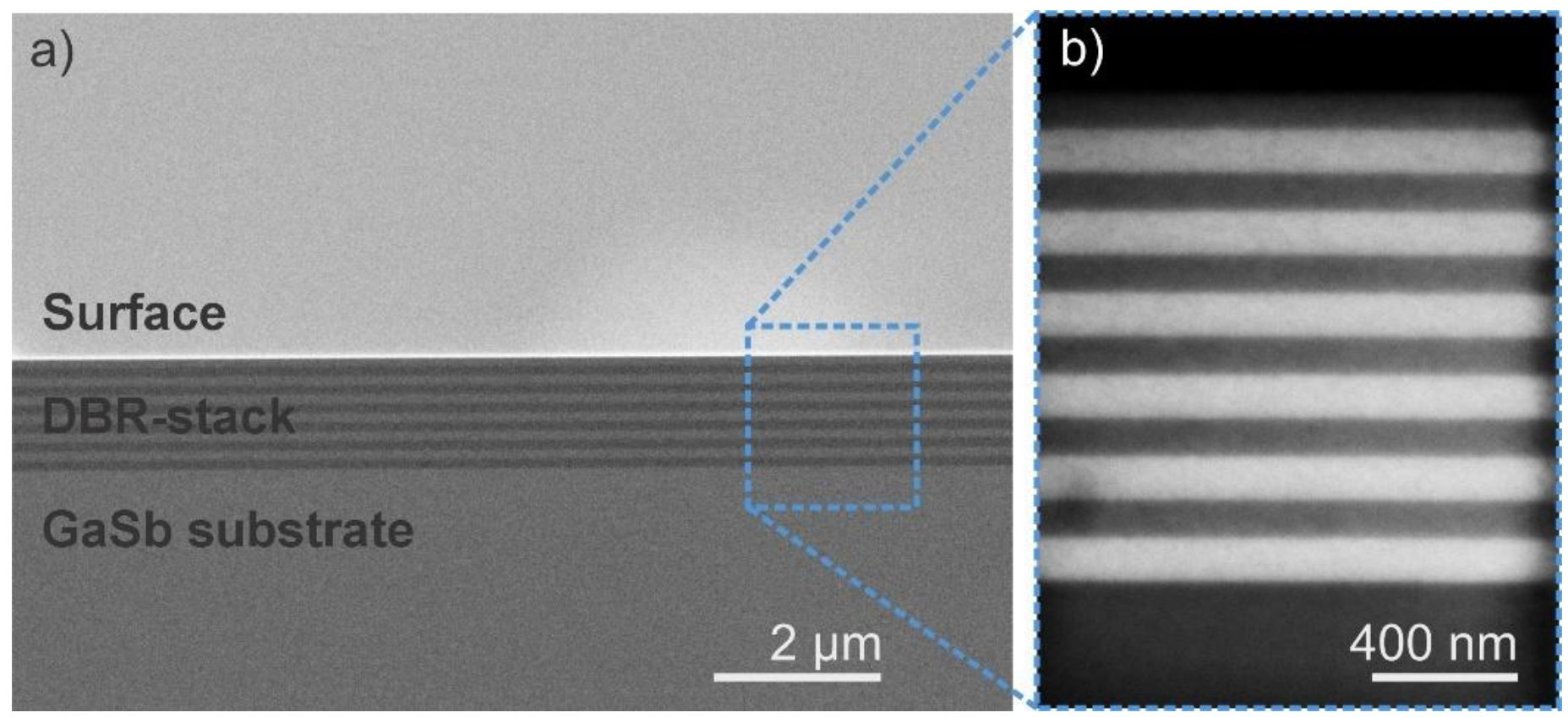

Fig. 2. Cross-sectional SEM overview (a) and a high-magnification view (b) secondary electron images of sample A.

## 3. Methodology

For establishing the temperature dependence of refractive index for Sb-based alloys we combine reflectivity measurements of abovementioned photonic structures with numerical modelling of respective spectra using TMM [18].

In this study we measured reflectivity of the investigated samples (shown in Fig. 1) with spectrally broad light source (halogen lamp). The light beam is mechanically modulated with the frequency of 286 Hz which is also the reference frequency for the nanovoltmeter with a lock-in amplifier. The samples were mounted in a closed-cycle He refrigerator with nominal base temperature of 10 K. Selected temperature was achieved and stabilized using temperature sensor and a heater mounted on the cold finger and a PID temperature controller. For data collection we use a Czerny-Turner monochromator of 30 cm focal length and InGaAs photodiode connected to the nanovoltmeter. To eliminate the spectral characteristics of the experimental setup, a reference reflection spectrum of a gold mirror in the identical conditions was taken. Spectra collected from the samples divided by the reference spectrum are further analyzed and presented within the manuscript. This procedure is based on the fact that the reflectance of gold in the spectral range of interest is almost constant and close to 100%.

In the next step, the reflectivity spectra were fitted with results of the calculations performed using TMM self-implemented in Python. The TMM approach is based on consideration of the amplitudes of electric field during the propagation of the electromagnetic wave through the structure. It includes the information of reflectance and transmission at each of the interfaces, thicknesses and refractive indices of each layer. With matrix notation, the light propagation through the whole structure might be represented by the 2D array called transfer matrix. Dividing and squaring the appropriate matrix elements may provide information about the reflection from the entire structure [20].

To determine the refractive indices of the materials, we applied a recursive algorithm proposed by A. Zielińska et al. [18]. The main fitting parameters are refractive indices of the materials involved. Their initial dispersions (for reflectivity spectra taken at 300 K) are calculated by formulas for AlGaAsSb alloys in room temperature, proposed by C. Alibert et al. [21]. Dispersion relations are shown in Fig. 3. At 1550 nm the calculated values of refractive indices are $n_{Al_{0.32}GaAsSb} = 3.699$ and $n_{Al_{0.9}GaAsSb} = 3.263$. To achieve the best possible fit to the experimental data and to consider uncertainties of the SEM measurements, we first optimized the thicknesses of the alternating layers (separately for each measurement series) based on spectra acquired at room temperature. The resulting thicknesses of the low-Al and high-Al-content layers for each series are as follows: 1) 102.9 nm and 119 nm, 2) 102.9 nm and 119.5 nm, and 3) 102.9 nm and 118.8 nm. In further calculations we use average thicknesses: 102.9 nm and 119.1 nm for low and high-Al-content layers respectively. There is no physical reason for the differences, so they can be treated as a measure of accuracy of their determination via the fitting procedure. The initial values of refractive indices taken from the literature are also varied for room temperature to obtain the best fit to experimental data. Obtained values are within the range reported in the literature [16]. Their change with temperature is then determined by minimizing function taking the spectrally integrated difference between two normalized spectra (measured and calculated) and is optimized in the range of 1350-1800 nm, where stopband occurs. The absolute value of the reflectance is more prone to details of the setup adjustment (probing beam size and angular adjustment required after change between the actual and reference sample) and it would also make it difficult to compare the results between different measurements series. Therefore, we decided to use normalized reflectance and focus on the shape of the spectrum to not introduce additional source of deviation in the obtained results. Normalization of the experimental reflectance spectra causes the model to be overparametrized, therefore the extracted refractive index of the GaSb is not reliable as its value scatter to compensate lack of information on the absolute value of the maximal reflectance for each experimental curve. In the case of the cavity structure, it is the same case with AlGaSb cavity layer. However, for the materials constituting the DBR, a requirement that refractive index dependence on temperature should be smooth and taking the refractive index at a given temperature as an initial value for the fitting procedure at the following temperature, is imposed. Therefore, at room temperature we used GaSb and $Al_{0.32}GaSb$ dispersion based on Ferrini et al. [11]. From the highest to the lowest temperature, spectra are fitted with initial values of refractive indices from previous reflectivity spectrum (at higher temperature). The scheme of fitting algorithm used to determine refractive indices at each temperature is shown in Fig. 4. In our calculations we neglect the thermal expansion of the materials which has previously been proven to have negligible impact on the results [18].

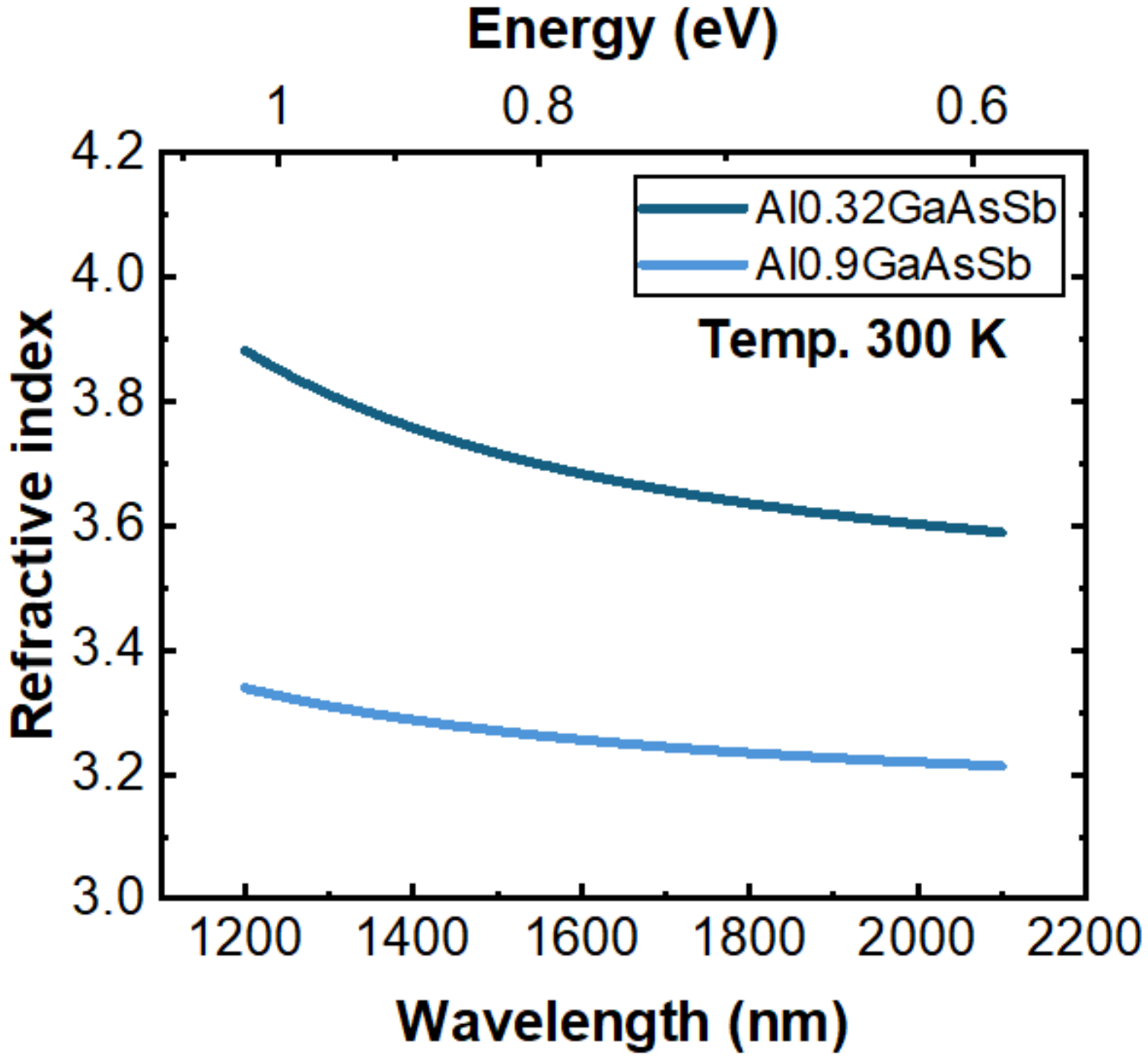

Fig. 3. Calculated dispersion relations for AlAs-GaSb alloys in room temperature, based on formulas proposed by C. Alibert et al. [21].

The procedure is performed on results obtained for sample A (DBR only) and determined refractive indices are further used to model the reflectivity spectra of sample B (optical cavity) for verification.

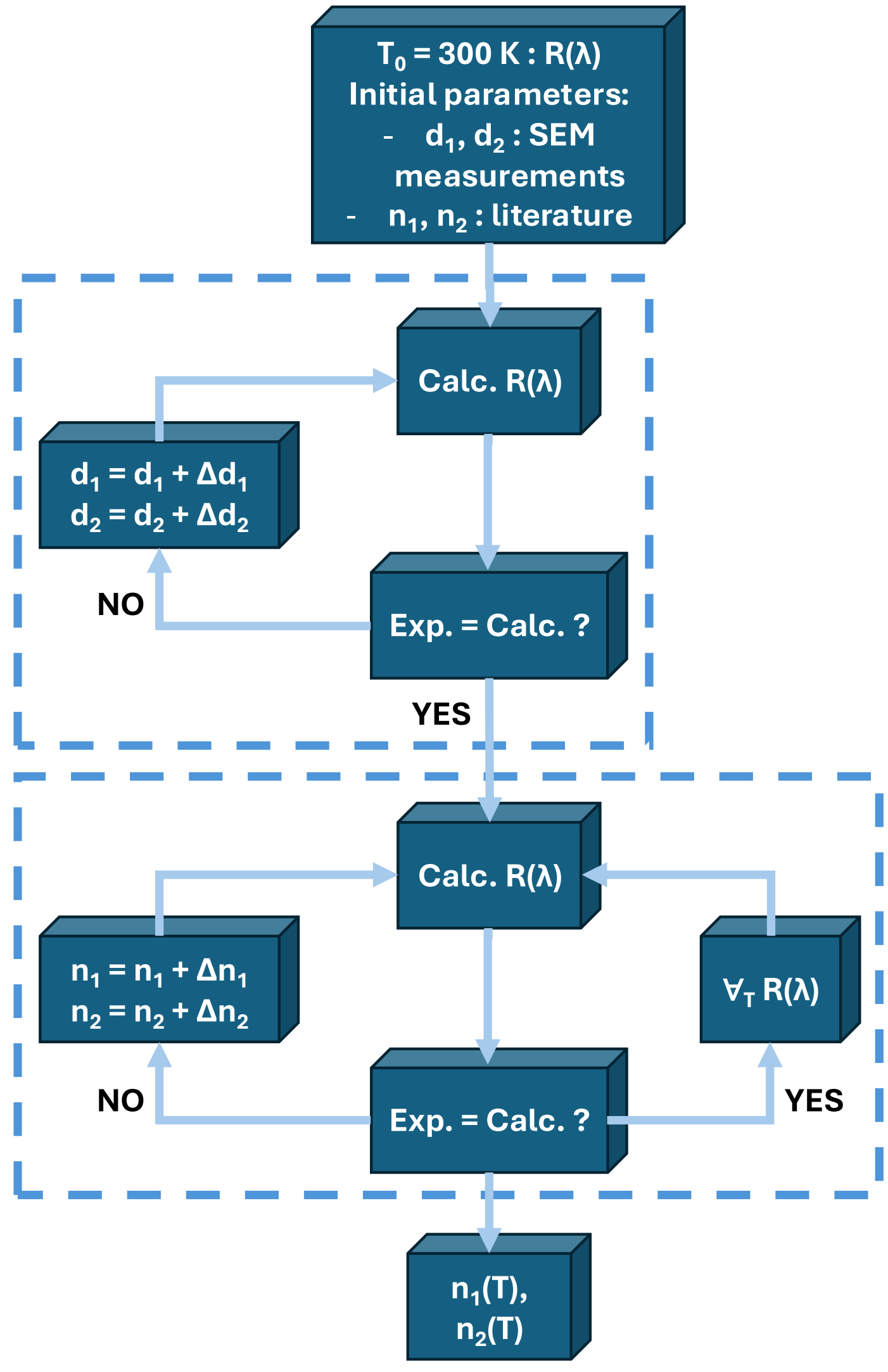

Fig. 4. The scheme showing the algorithm to find refractive index change with temperature.

## 4. Results

Firstly, the reflectivity spectrum of sample A was taken. The measurements were conducted in three independent series within temperature ranges of 11.5 K–300 K, 12 K–295.3 K and 150 K–300 K, to better estimate the accuracy of utilized approach. The last measurement series was carried out in the range of the most prominent differences between the two initial measurement series. Each obtained data set is included in **Data File 1**, **Data File 2** and **Data File 3** respectively. The resulting normalized reflectance spectra from the 1st series are shown in Fig. 5. With temperature the central wavelength of the stopband shifts between 1525 nm (at 11.5 K) and 1550 nm (at 300 K), which is rather typical value for different III-V material systems and can be traced back to variation in the refractive indices of the two materials [22, 18]. For the same reason the full width at half maximum (FWHM) of the stopband changes from 235 nm (11.5 K) and 255 nm (300 K).

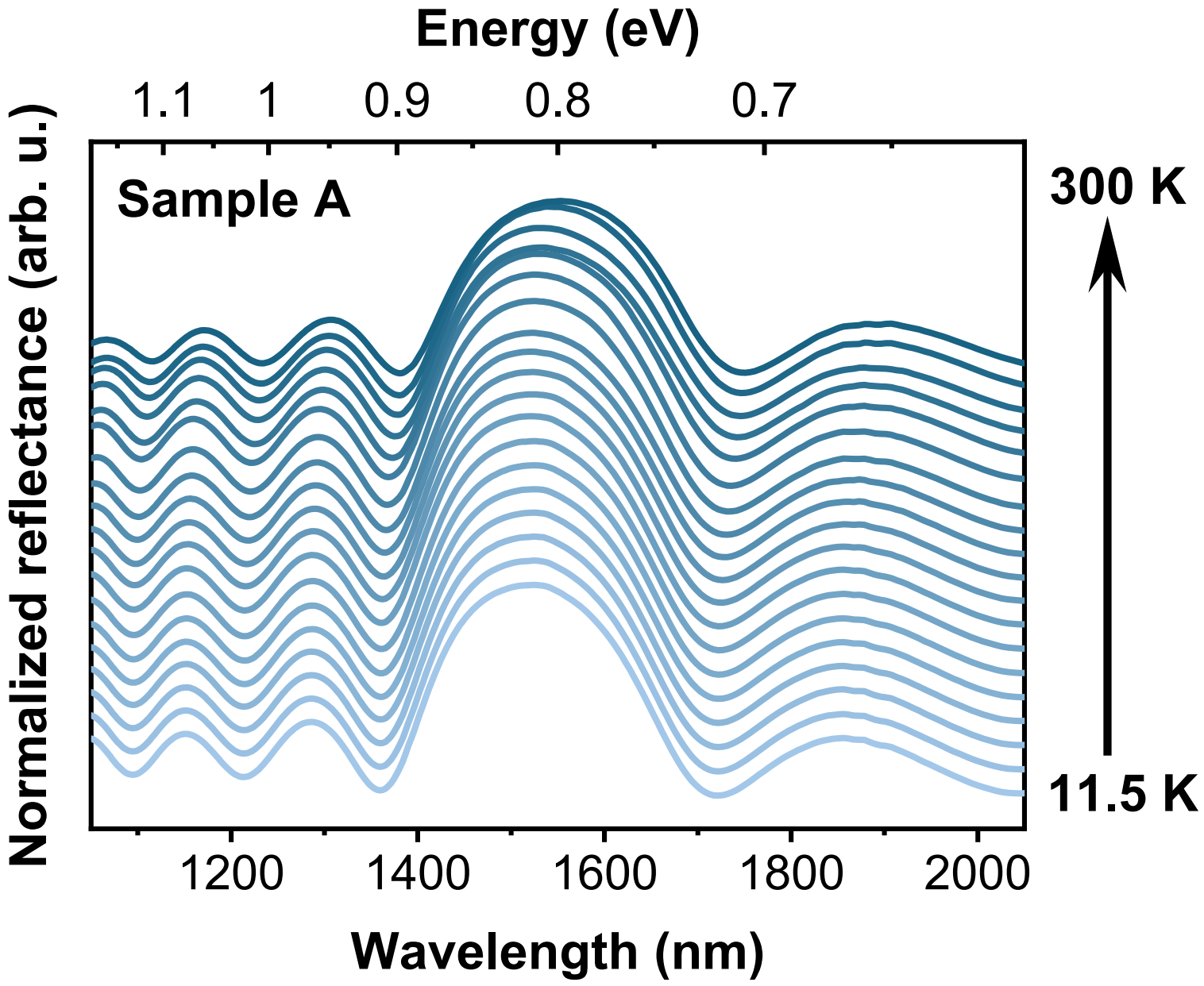

Fig. 5. Normalized reflectivity spectra of Sample A measured in a range of temperatures 11.5-300 K (referred to in the test as 1st measurement series). See **Data File 1** for underlying values.

To obtain information about refractive index of each layer type the data was fitted using results of TMM calculations. From the spectra taken at the highest temperature to the lowest, with fixed thicknesses of layers, the refractive indices of materials were adjusted to acquire the best possible fit. An example of measured data and fitted curves are shown below in Fig. 6. As can be seen the spectral range of the stopband and the first interference minima are very well-reproduced. The discrepancies appear for both spectral extremes of the measured spectra. In both cases it could be caused by lower signal to noise ratio in the measurements (strong decrease of the detector sensitivity for long wavelength range and low light level of the halogen lamp combined with low transmission of the monochromator for grating blazed at 1600 nm used in the measurements). To not introduce additional variation in the results we decided to measure the whole spectrum using the same experimental conditions, instead of optimizing them for each spectral range separately. We take this into account in our analysis by narrowing the spectral range for which the optimization procedure is performed. Once this range and room temperature dispersion is determined, the dispersion for different temperatures can be estimated assuming that its character stays constant but shifts vertically with temperature [23]. This however requires additional, relatively strong assumption, therefore, we would like to avoid it and instead focus on determination of temperature dependence of the refractive index for application-relevant centre of the telecommunication C-band. Next step was to extract the data of the refractive index change with the temperature. Procedure described above was repeated for all temperatures and measurement series. The resulting dependencies are presented in Fig. 7. The scatter points in color correspond to the data obtained from the TMM fitting for each measurement series with different temperature ranges. Black diamond points are mean values of the independent measurement sets, and solid lines are fitted curves. Error bars were calculated as standard error of mean value. In both cases the refractive index decreases with temperature, reaching low temperature (11.5 K) value of $3.6199 \pm 0.0013$ and $3.2452 \pm 0.0040$ for the low- and high-Al content alloy, respectively. The refractive index change over the 11.5-300 K temperature range differs between the two materials, and it is larger for the low-Al content alloy – $0.073$ compared to the high-Al one – $0.035$. Expressed as a relative change, the refractive index decreases by approximately 2% for the low-Al content alloy and 1% for the high-Al content alloy over the investigated temperature range. This difference motivates a detailed analysis of the temperature dependence of the refractive indices for both materials. DBR optimization based solely on the stopband shift with temperature, in the case of nonuniformly changing refractive indices, will not satisfy the constructive interference condition required for proper DBR performance. Therefore, determining the refractive indices of both materials is essential for accurate DBR optimization. The mean values were fitted with function determined by Taylor series expansion of refractive index around $T_0 = 300\ K$:

$$\frac{\partial n(\lambda,T)}{\partial T} = C_1(\lambda) + C_2(T - T_0) + C_3(T - T_0)^2, \tag{1}$$

giving the relations describing refractive index dependence on temperature at 1550 nm for studied materials:

$$n_{Al_{0.9}GaAsSb}(1550\ nm, T) = 3.2806 - 1.82 \cdot 10^{-4}(T - 300\ K) - 3.05 \cdot 10^{-6}(T - 300\ K)^2 - 6.92 \cdot 10^{-9}(T - 300\ K)^3 \tag{2}$$

$$n_{Al_{0.32}GaAsSb}(1550\ nm, T) = 3.6926 + 2.37 \cdot 10^{-4}(T - 300\ K) - 1.28 \cdot 10^{-6}(T - 300\ K)^2 - 4.26 \cdot 10^{-9}(T - 300\ K)^3 \tag{3}$$

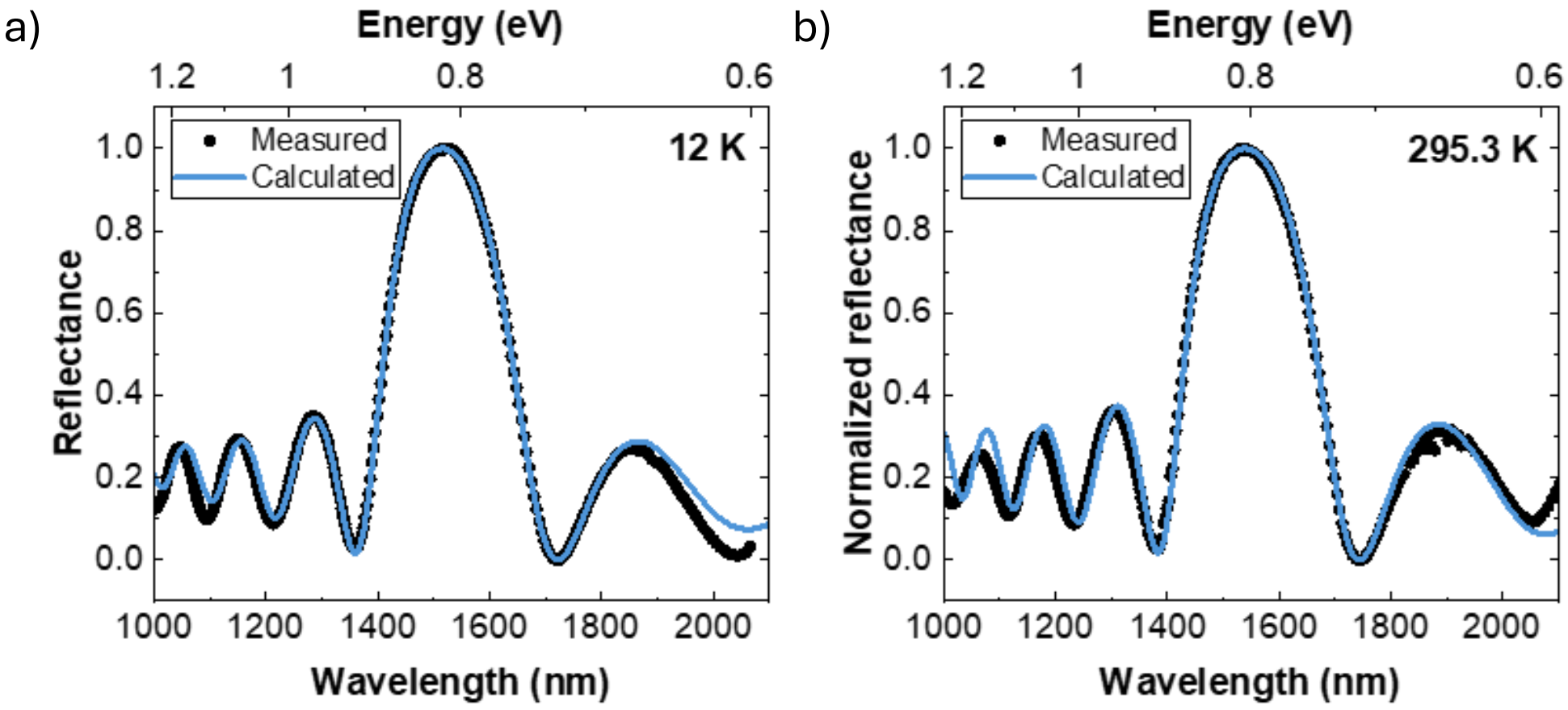

Fig. 6. Normalized reflectivity spectra of Sample A (a) at temperature of 12 K and (b) 295.3 K. Black data points represent measured data, blue solid lines are best fit to experimental data calculated using TMM.

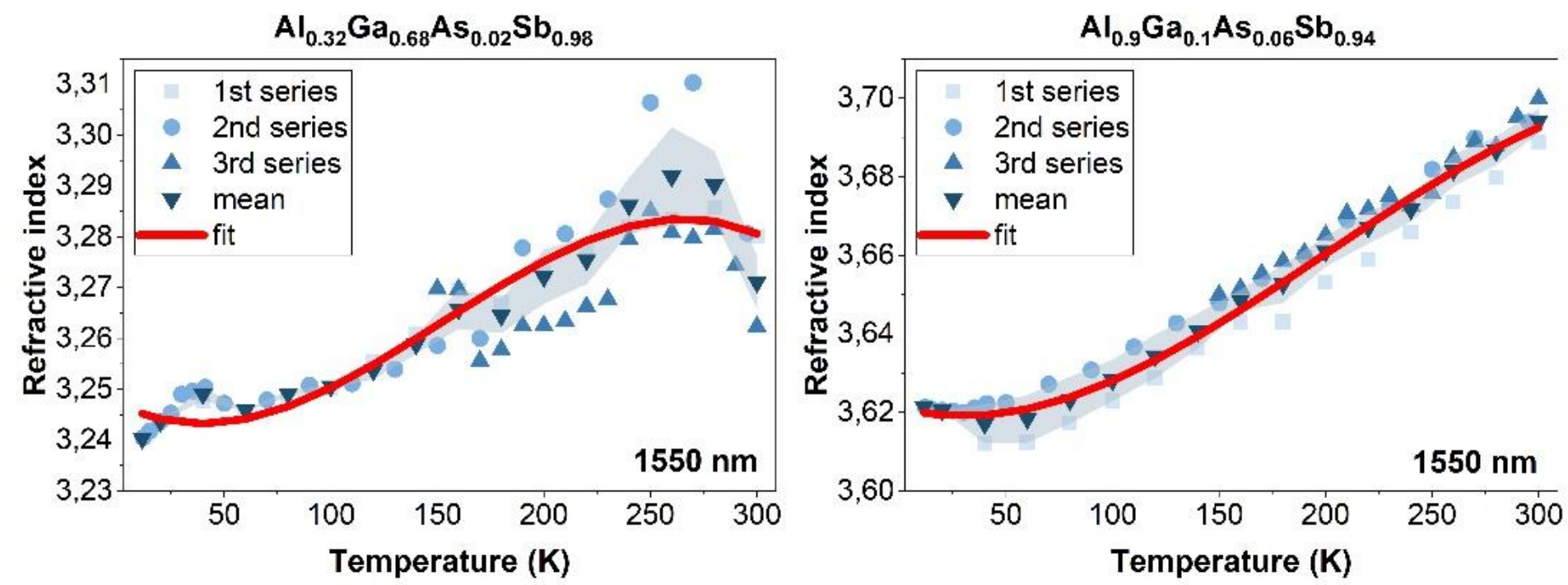

Fig. 7. Refractive index dependencies on temperature at 1550 nm for two studied materials: a) $Al_{0.32}Ga_{0.68}As_{0.02}Sb_{0.98}$ and b) $Al_{0.9}Ga_{0.1}As_{0.06}Sb_{0.94}$. The colored data points correspond to three reflectivity measurement series and black diamonds are their mean value calculated for each temperature with error area representing error of the mean value. The solid red line is the fit to the mean value temperature dependence with formula (2) and (3), respectively.

The results for $Al_{0.32}Ga_{0.68}As_{0.02}Sb_{0.98}$ are well reproduced by the second order Taylor polynomial, the high Al-content alloy might require more complex fitting function, but taking into account the experimental accuracy, lack of microscopic model justifying a specific functional dependence and the fact that the two alloys differ only in the exact content of the contributing atoms, it would not be reasonable to employ arbitrary fitting function to obtain better agreement. What is most important is the low-temperature refractive index value and reproducing the overall trend of the experimental data. Additionally, noteworthy is the fact that the greatest variation in obtained data is for the higher temperature range (approximately between 170 K and 300 K). Each series was taken at different measurement sessions, and therefore the sample was heated and cooled multiple times. In high-Al content alloy, the oxidation of aluminum may have contributed to the differences in measurements. The dispersion relation of refractive index in low temperature (12 K) is shown in Fig. 8. The shape of this curve follows Aliberti et al., but is shifted vertically, so the refractive index value at 12 K is as obtained by us.

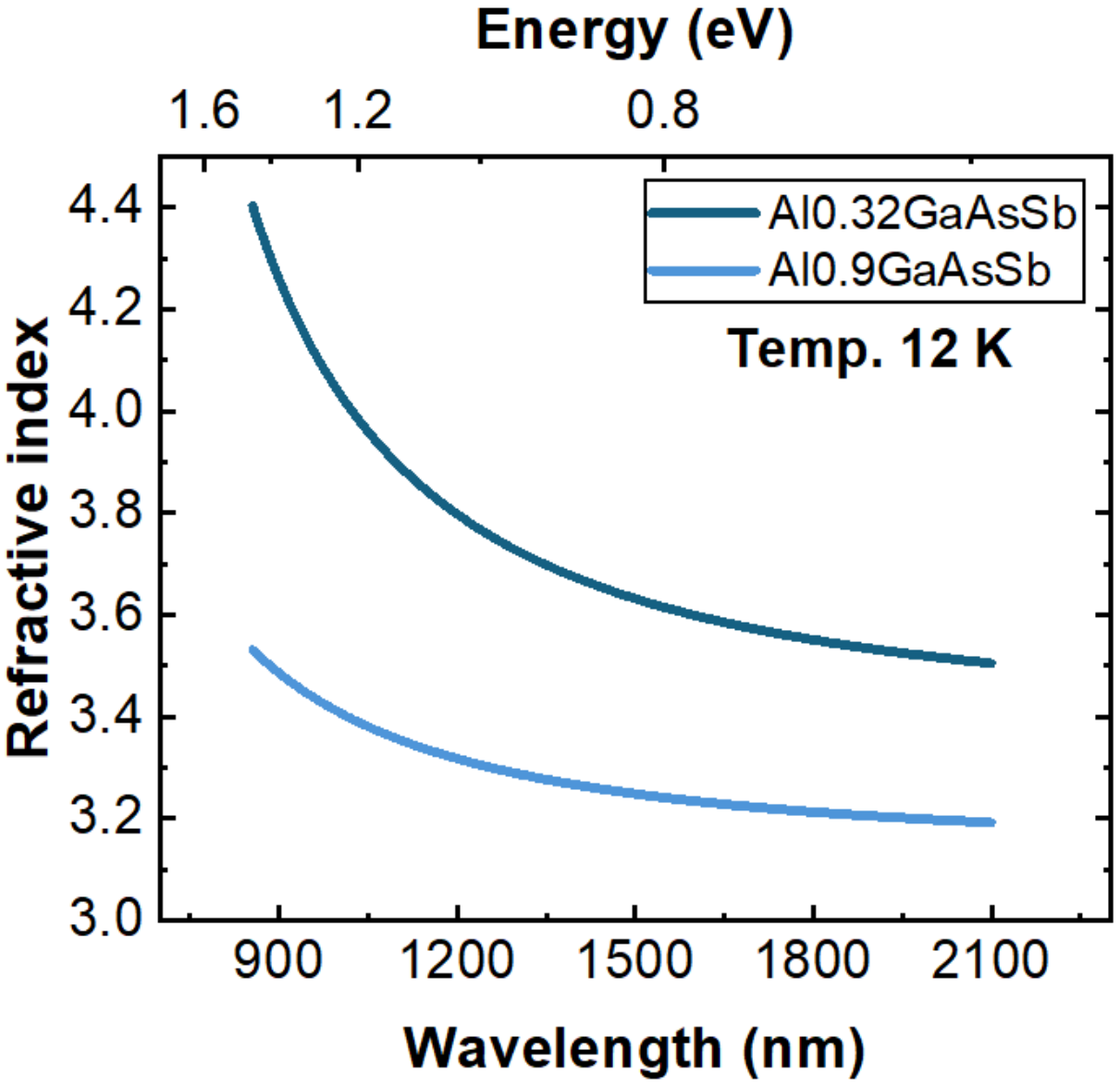

Fig. 8. Calculated dispersion relations for AlGaAsSb in 12 K.

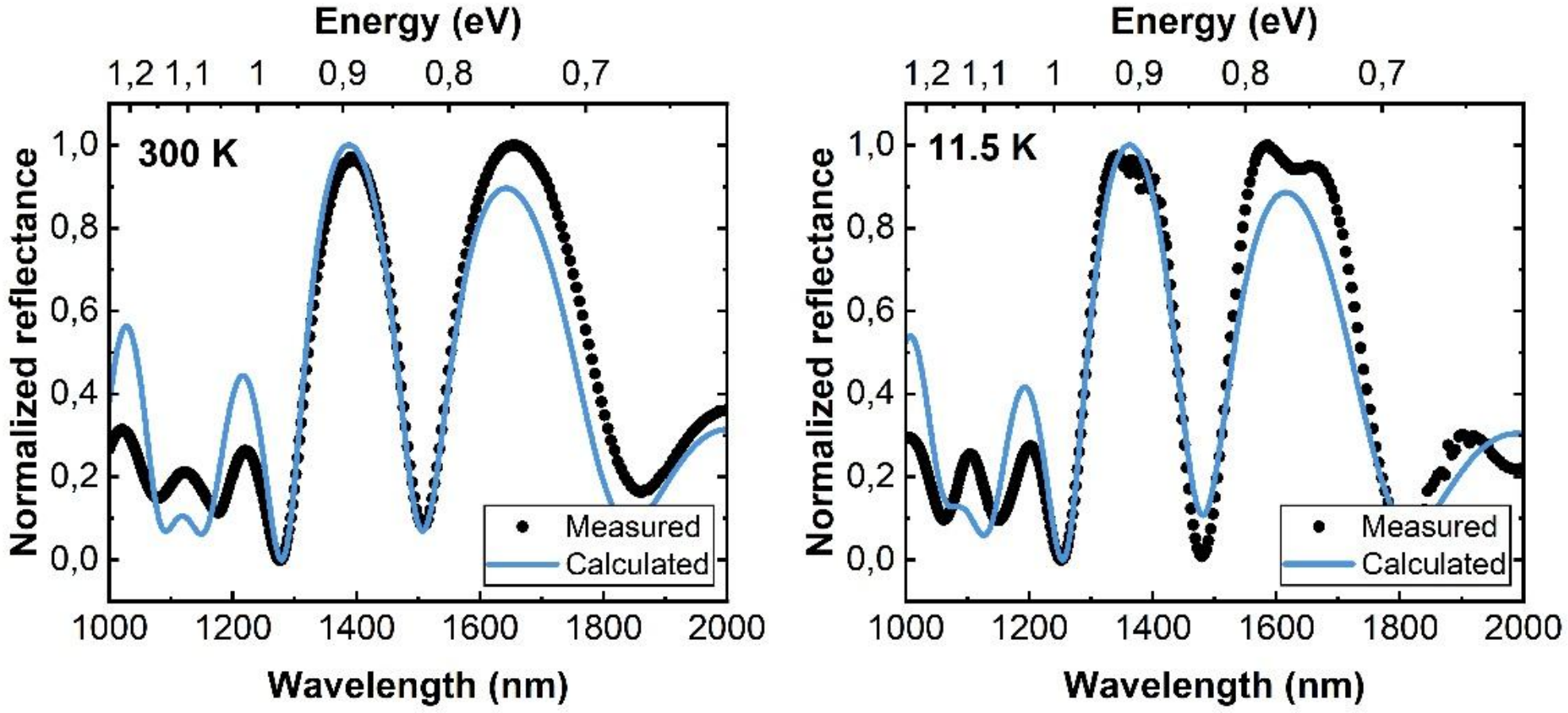

Fig. 9. Reflectivity spectra of Sample B took at a) 300 K, and b) 11.5 K fitted with determined dispersion relations. Black dots are experimental data, and blue lines are TMM fits with fixed thicknesses of layers and refractive index dispersions. See **Data File 4** for underlying measured values.

Lastly, to verify determined dispersion of refractive index of AlAsGaSb alloys of interest, we used it to fit Sample B reflectivity spectrum. Results for boundary temperatures (300 K and 11.5 K) are shown in Fig. 9. To obtain the best fit respectively to cavity mode position, we changed the thicknesses of the layers within uncertainty of SEM measurements. Within full temperature range we used values 100.4 nm for the high-Al content layer and 110.5 nm for the low-Al content layer. In both graphs calculated spectra qualitatively fit well to measured data in the spectral range of cavity mode and first interference maxima providing accurate prediction of the most important optical features.

## 5. Conclusions

In this work we determined the temperature dependence of refractive index for application relevant AlGaAsSb alloys lattice-matched to GaSb providing high refractive index contrast crucial for design of photonic structures. The spectral range of interest was a telecom C-band and the covered temperature range – 10-300 K. For both materials the refractive index decreases with temperature, but the change is faster in the case of low Al-content alloy, so the refractive index contrast also decreases. Utilized approach combining reflectivity measurements and TMM calculations allowed identifying low temperature (11.5 K) value of refractive index of $3.6199 \pm 0.0013$ and $3.2452 \pm 0.0040$ for 32% and 90% Al content, respectively. The inaccuracy of the method based on fitting precision and multiple measurements does not exceed 1% enabling precise design of GaSb-based photonic structures.

**Acknowledgment.** This work was financed by FiGAnti project funded within the QuantERA II Programme that has received funding from the European Union's Horizon 2020 research and innovation programme under Grant Agreement No101017733 and National Science Centre Poland – project 2023/05/Y/ST3/00125. Authors acknowledge financial support from Finnish Research Council project CryoLight (Decision No. 357351), Business Finland Rise to Challenge project TeleQuant (1835/31/2025). This work made use of Tampere Microscopy Center facilities at Tampere University.

**Data availability.** Experimental data underlying the results presented in this paper are available in Data Files 1-4.